\documentclass[aps,superscriptaddress,amssymb,showpacs,prb,twocolumn]{revtex4}
\usepackage{graphicx}
\newcommand{\rem}[1]{} % Per commentare senza togliere davvero e senza
\begin{document}

\title{Quantum annealing of an Ising spin-glass by Green's function Monte Carlo}

\author{Lorenzo Stella}
\affiliation{International School for Advanced Studies (SISSA), and INFM
Democritos National Simulation Center, Via Beirut 2-4, I-34014 Trieste, Italy}
\author{Giuseppe E. Santoro}
\affiliation{International School for Advanced Studies (SISSA), and INFM
Democritos National Simulation Center, Via Beirut 2-4, I-34014 Trieste, Italy}
\affiliation{International Centre for Theoretical Physics
(ICTP), P.O.Box 586, I-34014 Trieste, Italy}

\date{\today}

\begin{abstract}
We present an implementation of Quantum Annealing (QA) via lattice Green's function Monte Carlo (GFMC),
focusing on its application to the Ising spin-glass in transverse field.
In particular, we study whether or not such method is more effective than the
Path-Integral Monte Carlo (PIMC) based QA, as well as classical simulated
annealing (CA), previously tested on the same optimization problem.
We identify the issue of importance sampling, i.e., the necessity of possessing reasonably
good (variational) trial wavefunctions, as the key point of the algorithm.
We have considered two possible classes of trial wavefunctions, a mean-field single-site
one --- whose optimization is however a very difficult task --- and a Boltzmann-like choice. 
We performed GFMC-QA runs using such a Boltzmann-like trial wavefunction, finding
results for the residual energies that are qualitatively similar to those of CA
(but at a much larger computational cost), and definitely worse than PIMC-QA.
We conclude that, at present, without a serious effort in constructing reliable importance
sampling variational wavefunctions for a quantum glass, GFMC-QA is not a true competitor of PIMC-QA.
\end{abstract}

\maketitle

%------------------------------------------------------------------------------
\section{Introduction}\label{Intro:sec}
%------------------------------------------------------------------------------
 
Quantum annealing (QA)\cite{Das_Chakrabarti:book} is based on the idea of searching for the ground state of some 
classical Hamiltonian by adiabatically switching-off an appropriate source of quantum fluctuations,
in much the same way as temperature would do in thermal annealing.
This approach is also known as {\em adiabatic Quantum Computation}, in the Quantum Computing 
community \cite{Farhi_QP00:preprint}.
For recent reviews of the QA field, see Refs.~\onlinecite{Das_Chakrabarti:book,Santoro_JPA:review}.

A very popular QA approach is based on an imaginary-time Quantum Monte Carlo (QMC) implementation, i.e.,
the Path-Integral Monte Carlo (PIMC) approach.
A certain success in the application of PIMC-QA has been obtained in most of the cases studied: 
The folding of off-lattice polymer models,\cite{Berne1:article,Berne2:article} the random Ising model 
\cite{Santoro:science,Martonak:ising} and the random-field Ising model ground state search \cite{Sarjala_JSTAT06},
Lennard-Jones clusters optimization,
\cite{Berne3:article,Gregor-Car:article} and the traveling salesman problem.\cite{Martonak:TSP} 
Nevertheless, a counterexample exists,\cite{demian_QA:article} where PIMC-QA performs definitely 
{\em worse} than simple CA: The 3-Boolean-Satisfiability (3-SAT) problem,\cite{Papadimitriou_small:book} 
which is a prototype of a large class of hard combinatorial optimization problem 
(the so-called nondeterministic polynomial- ({\bf NP})- complete class, see Ref.~\onlinecite{Garey_Johnson:book}).

Generally speaking, the PIMC-QA failure depends on the properties of the ``landscape'' of the problem at hand
%(about which very little is known \cite{TSP_landscape:article})
or even to a bad performance of the implementation.
In order to understand its features in details, in a recent paper \cite{Stella_PRB06} 
we have studied the PIMC-QA performance focusing our attention on a simple, but highly instructive, 
toy-problem: The double-well potential.
There we learned a few potential dangers of the PIMC-QA method, in particular: 
i) The unavoidably finite temperature $T$ used in the simulation, which provides a thermal lower limit to the 
average residual energies attained by the algorithm.
ii) The possible severe difficulties (ergodicity breaking) in sampling
the PIMC action close to a Landau-Zener crossing of ground state levels.

We propose here to investigate a different Quantum Monte Carlo (QMC) based QA algorithm, as
an alternative to PIMC-QA.
A very natural choice is provided by the {\em Green's Function Monte Carlo} (GFMC). 
GFMC is different from PIMC since it can directly sample the ground-state of a quantum Hamiltonian, 
avoiding, in principle, one of the PIMC drawback, i.e., the finite temperature $T$.
However, contrary to PIMC, GFMC demands the knowledge of good variational wavefunctions to implement
what is called the {\em importance sampling}. We will appreciate soon how serious a drawback can
this be.
 
A natural benchmark for a test of this new GFMC-QA algorithm is provided
by the random Ising model, a challenging optimization problem already addressed 
through PIMC-QA, as well as standard classical simulated annealing (CA), 
in the recent past.\cite{Santoro:science,Martonak:ising}
The Hamiltonian of random Ising model in transverse field is:
\begin{equation} \label{Ising_glass_hamiltonian:eqn}
H(\Gamma) = -\sum_{\langle i,j\rangle}\, J_{i,j}\, \sigma^z_{i}\sigma^z_{j} 
- \Gamma\,\sum_{i}\,\sigma^x_{i} = H_{cl}+H_{kin}\;,
\end{equation}
where $\sum_{\langle i,j\rangle}$ indicates a sum over nearest-neighbors, 
$J_{i,j}$ are random nearest-neighbor Ising coupling constants, and $\sigma^z_{i},\sigma^x_{i}$ 
are Pauli's matrices on lattice site $i$.
If we denote by $\{ S_{i}\}$ a generic configuration in the Hilbert space 
(where $S_{i}=\pm 1$ are the eigenvalues of $\sigma^{z}_{i}$ matrix), 
the classical function we want to minimize is just given by the first term in 
Eq.~\ref{Ising_glass_hamiltonian:eqn}, 
i.e., $E_{cl}(\{S_i\})=\langle\{ S_{i}\}|H_{cl}|\{ S_{i}\}\rangle$
--- the Ising glass energy itself ---, which plays the role of {\em potential energy}.
The second term in Eq.~\ref{Ising_glass_hamiltonian:eqn}, 
$H_{kin}=-\Gamma\,\sum_{i}\,\sigma^x_{i}$, is the needed source of quantum fluctuation, 
which plays therefore the role of a {\em kinetic energy}.
More in detail, we will concentrate our efforts on a problem instance which has been analyzed extensively
in Ref.~\onlinecite{Santoro:science,Martonak:ising}:
it is a two-dimensional (2D) instance, on an $L\times L$ square lattice with $L=80$, and 
the couplings $J_{i,j}$ drawn from a flat distribution in $[-2,2]$.

Since we want to employ Eq.~\ref{Ising_glass_hamiltonian:eqn} as the Hamiltonian of our QA dynamics,
the transverse field $\Gamma$ represents the annealing parameter of the system; 
the goal is to follow the time-dependent dynamics with a $\Gamma(t)$ which starts from 
very large values, and is ramped down to zero in a certain annealing time $\tau$. 
We emphasize that such a transverse field term is not just a theoretical concept. Indeed, 
the whole field of QA was strongly revived by experimental results 
on the disordered Ising ferromagnet LiHo$_{0.44}$Y$_{0.56}$F$_4$,
where the transverse external magnetic field $\Gamma$ was actually applied to the system and manipulated in 
the laboratory, to perform a true {\em Quantum Annealing experiment}.\cite{Aeppli_QA1:article,Aeppli_QA2:article}

The rest of the paper is organized as follows.
In Sec.~\ref{gfmc_ideas:sec} we present the main ideas of a GFMC-based QA approach,
with a sketch of the main ingredients of the algorithm, for the benefit of the non-expert reader.
In Sec.\ref{Variational:sec} we present the results of our variational study of trial wavefunctions,
showing the inherent difficulties associated to the selection of good wavefunctions in a disordered
quantum system.
In Sec.\ref{static_GFMC:sec} we present the results of fixed-$\Gamma$ GFMC calculations.
In Sec.\ref{annealing_GFMC:sec} we discuss the GFMC-QA results, and compare them with previous
PIMC-QA and CA data on the same problem.
Finally, in Sec.~\ref{discussion:sec} we give some concluding remarks.
 
%-----------------------------------------------------------------------------------------------------
\section{Green's Function Monte Carlo Quantum Annealing: ideas}\label{gfmc_ideas:sec}
\begin{figure}[!ht]
\begin{center}
\includegraphics[width=6.5cm]{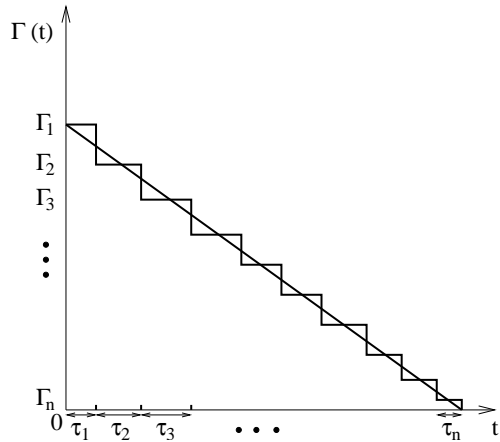}
\end{center}
\caption{Sketch of the annealing of the transverse field $\Gamma$, linearly
in a time $\tau=\tau_1+\tau_2+\cdots\tau_n$, or in a step-wise fashion.
}
\label{gamma:fig}
\end{figure}

%------------------------------------------------------------------------------
%
The ideal scope of a QA approach is to take some initial state $\psi(0)$ and let
it evolve, according to the Schr\"odinger dynamics associated 
to a time-dependent Hamiltonian $H(t)$ interpolating between an extreme quantum 
regime and the classical problem one is interested in: 
\begin{equation} \label{Schroedinger:eqn}
i\frac{d\psi}{dt} = H(t) \psi(t) \;,
\end{equation}
where we have set $\hbar=1$.
For instance, for the problem we have set to consider, we could take 
$H(t) = -\sum_{\langle i,j\rangle}\, J_{i,j}\, \sigma^z_{i}\sigma^z_{j} - \Gamma(t)\,\sum_{i}\,\sigma^x_{i}$,
where $\Gamma(t)$ is initially very large, and slowly decreased towards zero,
as illustrated in Fig.~\ref{gamma:fig}.
As argued in Ref.~\onlinecite{Stella_simple:article}, an imaginary-time
Schr\"odinger evolution would be, for optimization purposes, equally good, and most likely even superior to the 
standard real-time evolution. With this in mind, we can get rid of the $i$ in the time-derivative term in
Eq.~\ref{Schroedinger:eqn}, substituting it with the imaginary time prescription $-\partial_t$.
If we also imagine the gradual decrease of $\Gamma$ to be made step-wise, as sketched in Fig.~\ref{gamma:fig}, 
then the solution for $\psi(t)$ is obtained by repeated applications of an imaginary-time propagator 
\begin{equation}
\psi(\tau) = e^{-H(\Gamma_n)\tau_n} \cdots e^{-H(\Gamma_1)\tau_1} \psi(0) \;,
\end{equation}
where $\tau=\tau_1+\cdots\tau_n$ is the total annealing time, 
$\Gamma_1>\Gamma_2>\cdots >\Gamma_n \sim 0$ is a decreasing sequence of transverse fields,
and $H(\Gamma_i)$ is a shorthand for $H(t)$ with a value $\Gamma_i$ of the transverse field.
%
%-------------------------------------------------------------------------------------------
%-------------------------------------------------------------------------------------------
%
Each application of the imaginary-time propagator $e^{-H(\Gamma_i)\tau_i}$ effectively
tends to {\em filter out} the corresponding ground state of $H(\Gamma_i)$ from the state to which it is applied. 
In turn, $e^{-H(\Gamma_i)\tau_i}$ can be obtained as a repeated application of many infinitesimal 
propagators of the type $[1-\Delta t H(\Gamma_i)]$, i.e.,
\begin{equation} \label{gfmc_2:eqn}
\psi(t+\tau_i) = e^{-H(\Gamma_i)\tau_i} \psi(t) = \prod \left[ 1-\Delta t H(\Gamma_i) \right] \psi(t) \;.
\end{equation}
The Green's function Monte Carlo (GFMC) is just a stochastic technique which 
implements such a form of propagation.
More precisely, if we define, recursively,
\begin{equation} \label{power:gfmc}
\psi_{n+1}(x') = \sum_x \,G^{(\Gamma)}_{x',x} \psi_{n}(x)\;,
\end{equation}
where $\psi_n(x)=\langle x | \psi_n\rangle$, $|x\rangle$ being a shorthand for a
generic spin-configuration describing the Hilbert space of the problem, and
\begin{eqnarray} \label{Green_function:eqn}
G^{(\Gamma)}_{x',x} &=& \langle x' |G^{(\Gamma)}|x\rangle 
= \langle x'| 1 - \Delta t \,[H(\Gamma)-E_T] |x\rangle \nonumber \\
&=& (1+\Delta t\,E_T)\,\delta_{x',x} - \Delta t \langle x' |H(\Gamma)|x\rangle\;,
\end{eqnarray}
one can show that -- for large $n$ -- the iterated state $\psi_n$ converges
(apart from a normalization constant) to 
the ground-state $\psi_{GS}^{(\Gamma)}(x)$ of $H(\Gamma)$, if $\Delta t$ is chosen to
be suitably small.\cite{trivedi_ceperley:article}
($E_T$ is an estimate of the ground-state energy which allows to reduce the statistical fluctuations,  
see also Ref.~\onlinecite{Mitas:review}.)

The problem in Eq.~\ref{power:gfmc} looks superficially similar to an ordinary Markovian
Master equation, with a few very crucial differences:
i) The $\psi_n(x)$ are not {\em probabilities}, but {\em amplitudes};
%and as a consequence we do have (in general) that $\sum_x \psi_n(x) \neq 1$;
ii) The Green's function $G^{(\Gamma)}_{x',x}$ in Eq.~\ref{Green_function:eqn},
unlike the transition probability of a Master equation, is not necessarily made 
of non-negative elements, and is, in general, not column-normalized,
$\sum_{x'} G^{(\Gamma)}_{x',x} \ne 1$, unlike a Markov transition probability. 
In summary, the process underlying the iterated-power method is {\em not} a
properly defined Markov chain, and, therefore, it cannot be immediately simulated,
as it stands, with a Monte Carlo approach.

Problem ii) above can be quite serious: If some of the
matrix elements of $G^{(\Gamma)}_{x',x}$ are {\em negative}, no possible interpretation
of it as a ``transition probability'' is possible.
This is at the heart of the so-called {\em sign problem}\cite{Mitas:review} in Quantum Monte Carlo. 
In the following, we will assume that a choice of basis is possible in which
no sign problem exist, i.e. all matrix elements of $G$ are non-negative, $G_{x',x}\geq 0$.
This is certainly true for the Ising glass in a transverse field. More generally, since the
choice of the kinetic energy to be used in QA is at our disposal, it is wise to choose the signs of $H_{kin}$
in Eq.~\ref{Ising_glass_hamiltonian:eqn} such that no sign problem occurs.
Still, we miss the correct column-normalization,
\begin{equation} \label{normalization:eqn}
\sum_{x'} G^{(\Gamma)}_{x',x} = 1 + \Delta t\,E_T- \Delta t \sum_{x'} H_{x',x}
\stackrel{{def}}{=} b_x \ne 1 \;.
\end{equation}
A way out of this difficulty is to factorize $G^{(\Gamma)}$ in terms of a stochastic matrix $p_{x',x}$
-- by definition, a matrix with all positive elements $p_{x',x} \geq 0$, and with
the normalization condition $\sum_{x'} p_{x',x}=1$ for all columns --
times the scale factor $b_x$ defined above.
Indeed, with the previous definition (\ref{normalization:eqn}) of $b_x$, the matrix
\begin{equation} \label{defbx}
p_{x',x} = G^{(\Gamma)}_{x',x}/b_x  \;.
\end{equation}
is trivially positive and column normalized and, therefore,
it is a suitable transition matrix for a Markov chain in $x$-space.

The crucial idea is then to \emph{extend} the configuration space where the Markov process
is defined, adding to the $x$ a \emph{non-negative} weight factor (hereafter, the weight) $w$:
This extended configuration space will be labeled by $(x,w)$.
The pair $(x,w)$ is often called a \emph{walker}, because it will be the
basic entity in the Markov chain ``random walk''.
The weight part will take care of $b_x$, while $x$ will be taken care of by $p_{x',x}$.
More precisely, if $(x_n,w_n)$ indicates a walker at iteration time $n$,
in this extended configuration space, we set up the following Markov process:
\begin{eqnarray} \label{markovgf}
&& \mbox{a) generate $x_{n+1}=x'$ with probability $p_{x',x_n}$ }
\nonumber \\
&& \mbox{b) update the weight with $w_{n+1}=w_n b_x$ } \;.
\end{eqnarray}
In words: The walker performs a {\em random walk} in the Hilbert space $x$ of the system
and in the weight space $w$; such a random walk is composed of a standard Markov chain in
$x$-space, associated to the $p_{x',x}$, plus a multiplicative process
for the weight $w_n\to w_{n+1}=w_n b_x$.
By moving in this way, the walkers visits every point in the $(x,w)$-space with a probability 
$P_n(x_n,w_n)$ whose first moment can be shown to be proportional to the $\psi_n$, 
$\psi_n(x) \propto \int {\rm d}w_n\, w_n \, P_n(x,w_n)$.\cite{CalandraSorella:article}

Eq.~\ref{markovgf} is the basic version of a GFMC algorithm.\cite{trivedi_ceperley:article} 
In this form, however, the algorithm simply {\em does not work} in practice. 
The reason for this failure is not difficult to grasp. 
While $x_n\to x_{n+1}$ is a plain Markov process, the weight update
$w_n\to w_{n+1}=w_n \, b_{x_n}$ is a {\em multiplicative} process with random factors $b_{x_n}$,
which is prone to very large fluctuations:\cite{trivedi_ceperley:article,CalandraSorella:article}
$w_n$ might grow large, or become negligibly small, in just a few iterations,
and the whole algorithm would go wild, because error bars in the calculations of the averages
grow in an uncontrolled way.
The cure to this disease goes through the introduction of {\em many walkers} and through performing
occasional ``reconfigurations'' of their weights, via the so-called {\em branching}.\cite{Mitas:review}
In practice, one propagates simultaneously a set of $M$ walkers defined by weights $w_i$ and configurations
$x_i$, for $i=1,\cdots M$.
Before the variance of the weights $w_i$ becomes too large, one appropriately redefines the set
of walkers -- by reproducing some of them and deleting some others --
in such a way as to drop those with excessively small weight, and to generate copies of the 
more important ones.\cite{footnote10}

The analogy of such a many-walker GFMC with a genetic-like algorithm is worth noting. 
Each walker $(x,w)$ plays the role of an individual that propagates (mutates) increasing or decreasing
its {\em fitness} -- represented by the accumulated weight $w$, related to the wavefunction amplitude
$\psi(x)$. A mutation is here simply a step of the algorithm, which attempts a single spin-flip
of a certain site in the configuration: this is what the off-diagonal matrix elements 
$\langle x\prime|H|x\rangle$ do.
At certain times, branching occurs, which modifies the population of individuals by
favoring the survival of those with highest fitness (largest $w$). The only genetic feature that is missing
in the quantum mechanical case is the possibility of cross-breeding (mixing of genetic codes of
two configurations, to give rise to new configurations): this would correspond to {\em non-local moves}
which change the configurations in a global way.

The final, important, ingredient that makes the algorithm work is the 
so-called {\em importance sampling}.\cite{Kalos:book} It can be seen, in the genetic analogy 
proposed before, as a way of proposing mutations (single spin flips) that
instead of being equally probable, with a matrix element $\Gamma$, are biased
by a function which guides the system towards the most representative configurations.
More precisely, suppose we have a reasonable guess of the ground state $\psi_{GS}(x)$ 
in the form of some nodeless wavefunction $\psi_T(x)$, known as {\em trial (or guiding) wavefunction}.
It is then enough to substitute $G$ with the so-called {\em importance sampling} Green's function: %!!!
\begin{equation} \label{importance}
\bar{G}_{x',x} = \psi_T(x') G_{x',x} \psi_T^{-1}(x) \;,
\end{equation}
which just rescales by an extra factor $\psi_T(x')/\psi_T(x)$ a transition from $x$ to $x'$,
thus favoring those transitions where $\psi_T(x')/\psi_T(x)$ is largest.
In general, $\bar G_{x',x}$ is not symmetric, but one can still apply to it
%if $\bar{G}_{x',x} \ge 0 $ for all $(x',x$), 
the same decomposition in (\ref{defbx}), defining the corresponding Markov chain (\ref{markovgf}) with:
\begin{eqnarray} \label{defnewbx}
p_{x',x} & =&  \bar{G}_{x',x} /\bar{b}_x  \nonumber  \\
\bar{b}_x
&=&\sum_{x'} \bar{G}_{x',x} = 1+\Delta t \,E_T -
\Delta t\,\frac{\sum_{x'}\psi_T(x')H_{x',x}}{\psi_T(x)}
\nonumber \\
&=&  1 - \Delta\,t\,\left( E_L(x) - E_T \right)\;,
\end{eqnarray}
where the local energy $E_L(x)$ is defined as:
\begin{equation}\label{local_energy:eqn}
E_{L}(x) = \frac{ \langle \psi_{T} | H |x \rangle }{ \langle \psi_{T}|x \rangle } \;.
\end{equation}
If the guessed trial wavefunction $\psi_T$
coincides with the actual ground-state wavefunction
$\psi_T(x)=\psi_{GS}(x)$, then $E_L(x)=E_{GS}$ is a constant, and
one can show that statistical fluctuations in the calculation vanish exactly.
This is the so-called \emph{zero variance property}.\cite{Mitas:review}
Therefore, by variationally improving the quality of the guiding function $\psi_T(x)$
one can substantially reduce the error bars in the calculation. %!!!

%-------------------------------------------------------------------------------------------
\begin{figure}[!ht]
\begin{center}
\includegraphics[width=8.5cm]{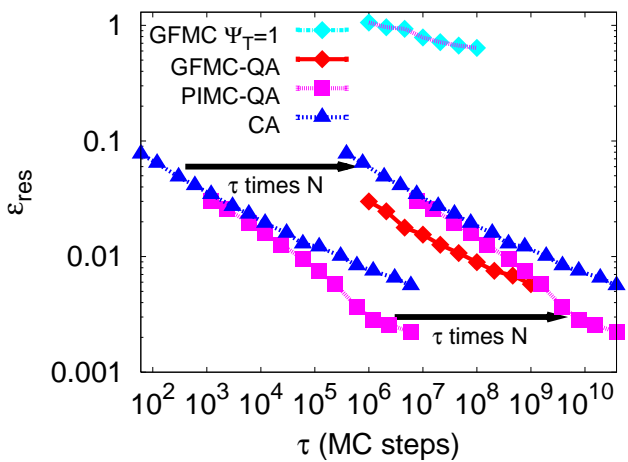}
\end{center}
\caption{(Color online) 
The average best residual energy obtained by GFMC-QA for the $80\times 80$ instance 
of the random Ising model studied in Refs.\onlinecite{Santoro:science,Martonak:ising}, 
versus the total annealing-time $\tau$. 
Upper rhombi: GFMC-QA results without importance sampling ($\psi_{T}=1$). 
Lower rhombi: GFMC-QA results with importance sampling performed by using the optimal 
trial wavefunction $\psi_{T}^{(\beta)}$ of Sec.~\ref{Variational:sec}. 
The GFMC time-unit is a single spin-flip, while CA and PIMC-QA Monte Carlo time units are 
sweeps of the entire lattice (see Ref.~\onlinecite{Martonak:ising}).
The transverse field is linearly reduced down to $10^{-4}$ in a total annealing-time $\tau$, 
starting from $\Gamma_0=2.5$. 
We used here $M=20$ walkers and performed branching at every MCS ($t_B=1$).
Previous results obtained by classical simulated annealing (CA) and by Path-Integral Monte Carlo 
quantum annealing (PIMC-QA) with $P=20$ Trotter slices \cite{Santoro:science,Martonak:ising} are 
shown for comparison.
%Results obtained for larger branching time (every 10 MCS, $t_B=10$) are compared to the
%present ones in Fig.~\ref{fig_glass_compare_ANN2_b0:fig}. 
}
\label{fig_annealing:fig}
\end{figure}
%-------------------------------------------------------------------------------------------
%
The fact that importance sampling is indeed a crucial ingredient is demonstrated, for our case,
by Fig.~\ref{fig_annealing:fig}, where we show the results 
obtained by GFMC annealing {\em without importance sampling} (top curve) compared
with the results obtained with importance sampling, which we will illustrate later on.
Quite evidently, the residual energy obtained without importance sampling is
terribly bad: any short classical simulated annealing would do better. 

Finding a reasonable trial function $\psi_T(x)$ for the problem at hand is therefore 
an essential part of a GFMC-QA application, and constitutes the delicate point 
of the whole algorithm. In the next section we will describe the choices of $\psi_T(x)$
we have tested for the Ising case, and the difficulties encountered.

%----------------------------------------------------------------------------------
\section{Variational wavefunctions for the Ising spin-glass}\label{Variational:sec}
%----------------------------------------------------------------------------------
%
Finding a good trial wavefunction for a random Ising model in a transverse field
is a highly non-trivial task. 
The first idea that comes to mind is a kind of ``mean-field'' wavefunction, 
made up of a product of single-site factors as:
\begin{equation} \label{mean_field_trial_function:eqn}
|\psi_{T}^{(MF)}\rangle = \prod_{i=1}^N\,\left(\frac{e^{+\frac{h_{i}}{2}}\,|\uparrow\rangle_{i}
+e^{-\frac{h_{i}}{2}}\,|\downarrow\rangle_{i}} {\sqrt{2\,\cosh(h_i)}} \right) \;,
\end{equation}
where $\{ h_i\}$, the local fields on each site $i$, are variational parameters to be optimized
for each given value of the transverse field $\Gamma$.
The optimization of the $\{h_i\}$ amounts to finding the minimum of the variational energy, 
$ E_{T}^{(MF)}=-\sum_{\langle i,j\rangle}J_{i,j}\,m_{i}\,m_{j} - \Gamma\sum_{i}\,\sqrt{1-m_{i}^2}$, 
where $m_i=\tanh(h_i)$ are the local magnetizations. 
The stationarity conditions required by the minimization, 
$\partial\,E_{T}^{(MF)}/\partial\,h_{i}=0$, read for each site $i$: 
\begin{equation} \label{minimization_MF:eqn}
-\left(1-m_{i}^2\right)\,\left( \sum_{j\in \mathcal{N}(i)}\,J_{i,j}\,m_{j}\right)
+\Gamma\,m_{i}\,\sqrt{1-m_{i}^2} = 0 \;,
\end{equation}
where $\mathcal{N}(i)$ indicates the set of nearest-neighbors of site $i$.
As it turns out, finding solutions of Eq.~(\ref{minimization_MF:eqn}) with optimal 
variational energies is simple only for large enough $\Gamma$, where the
quantum paramagnetic solution $h_i=m_i=0$ is found. 
%
%---------------------------------------------------------------------------------------------------
\begin{figure}[!ht]
\begin{center}
\includegraphics[width=8.5cm,angle=0]{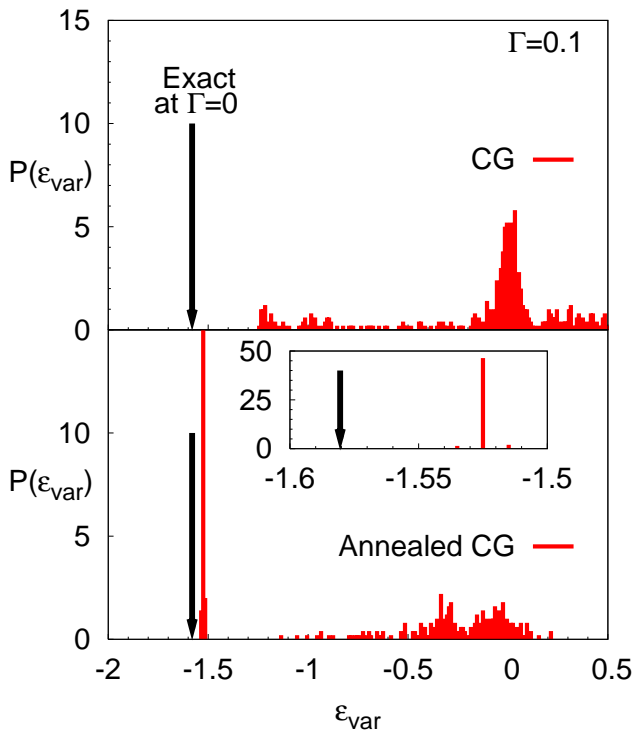}
\end{center}
\caption{(Color online) 
Distribution of the variational energy per spin
$\epsilon_{var}^{(MF)} = {\rm min}_{\{h_i\}} \langle \psi_{T}^{(MF)} |H|\psi_{T}^{(MF)} \rangle / N$ 
obtained by optimizing, with Conjugate Gradients, the local-fields $h_{i}$ in the trial wavefunction 
$|\psi_{T}^{(MF)}\rangle$ defined in Eq.~\protect\ref{mean_field_trial_function:eqn}.
These results refer to the case of a fixed $\Gamma=0.1$.
The upper panel has been obtained by repeated conjugate gradient (CG) minimizations,
while in the lower panel we have used an annealing CG scheme (see text).
In both cases, the initial local fields $h_i$ are randomly taken from the interval $(-0.5,+0.5)$. 
%The histograms are obtained by dividing the energy interval $[-2,0.5]$ into $250$ subintervals.
}
\label{fig_histo_CG:fig}
\end{figure}
%---------------------------------------------------------------------------------------------------
%
Such a solution, representing the $\Gamma=+\infty$ ground state with all spins aligned 
along the $+\hat{x}$ direction, survives down to some value $\Gamma_{cr}$ of the 
transverse field, below which non-trivial solutions of Eq.~\ref{minimization_MF:eqn} 
-- with non-vanishing local magnetizations $m_i\ne 0$ -- start to appear.
However, the number of those solutions (local minima) is large. 
In the low-$\Gamma$ region, our minimization problem is just the quantum counterpart of 
the well-known Weiss mean-field approach for the classical random Ising model,\cite{Binder-Young:review} 
which is known to run into difficulties in the classical glassy phase.
In a sense, minimizing $E_T^{(MF)}$ in the low-$\Gamma$ glassy phase is not much
simpler than finding the classical ground state of the problem for $\Gamma=0$: 
We transformed a minimization task in a discrete space of variables, $S_i=\pm 1$, into one 
where the variables are continuous, $m_i\in (-1,1)$, but the task itself is of comparable difficulty.
We anticipate that difficulties similar to the classical Weiss approach will plague our search for 
the minima of $E_T^{(MF)}$ in the low-$\Gamma$ phase. 

% RESULTS: MF function
Given the explicit analytic expression for the energy to be minimized, %Eq.~\ref{trial_energy_MF:eqn}, 
we made use of a standard Conjugate Gradients (CG) algorithm to find local minima solutions for the $\{h_i\}$, and 
the corresponding optimal value of the variational energy per spin, 
$\epsilon_{var}^{(MF)}={\rm min}_{\{h_i\}} E_{T}^{(MF)}(\{h_i\})/N$. 
However, as anticipated, the results obtained for $\epsilon_{var}^{(MF)}$ when $\Gamma<\Gamma_{cr}$ depend 
on the initial conditions of the algorithm, so that the only meaningful thing to do is to show 
{\em histograms} of $\epsilon_{var}^{(MF)}$, obtained by repeated --
and uncorrelated -- CG-minimization searches.
In Fig.~\ref{fig_histo_CG:fig} we report the results of
two different calculations for $\Gamma=0.1$,
which turns out to be in the low-$\Gamma$ glassy phase.
The top panel displays the histogram of $500$ repeated searches by CG-minimization, 
each search starting from initial $h_i$ which are randomly distributed in the interval $(-0.5,+0.5)$. 
The large solid arrow marks the location of the classical ground-state energy per spin, 
$\epsilon_{GS}=E_{GS}/N\simeq-1.5805167$, as a reference. 
In most of the attempts, we simply find minima with $\epsilon_{var}^{(MF)}\simeq 0$,
that have nothing to do with the classical GS which should dominate 
for such low $\Gamma$ value;
only rarely, we end up with local minima in the range $\epsilon_{var}^{(MF)}\simeq -1.215$, 
which is still quite far from $\epsilon_{GS}$.
Although, in principle, quantum effects due to the finite value of $\Gamma$ make 
$\epsilon_{T}^{(MF)}\neq\epsilon_{GS}$, the small $\Gamma$ value used ($\Gamma=0.1$) 
does not justify such large difference between the two quantities.
This is a clear suggestion that the bare CG-minimization approach applied to the 
single-site variational \emph{Ansatz} $E_T^{(MF)}$ is prone to get stuck 
in the high-energy paramagnetic phase.
Data (not shown, but monitored during the simulation)
on the average magnetization, the Edward-Anderson parameter,\cite{PMV:book} and the superposition
integral between the optimized variational state and the true ground-state, 
are also in agreement with this hypothesis.
To give further support to this conclusion, we show in the bottom panel of
Fig.~\ref{fig_histo_CG:fig} the results obtained by an improvement of 
the CG scheme, which is a sort of ``variational'' annealing. 
There we started from $\{h_i\}$ random distributed in $(-0.5,0.5)$
and we performed a first CG-minimization at $\Gamma_{0}=4$, a sufficiently high value
which provides an initial paramagnetic state. 
Then, the transverse field $\Gamma$ was repeatedly decreased in 10 steps
of $\Delta\,\Gamma = 0.39$ down to the desired value of $\Gamma=0.1$, and
at each step, the new minimum was searched via CG-minimization
starting from the optimal solution found in the previous step.
This ``variational annealing'' scheme provides results for $\epsilon_{T}^{(MF)}$ which are distributed 
much closer to the classical ground-state energy, $\epsilon_{GS}$, 
although we still obtain, in a non-negligible fraction of the attempts, paramagnetic solutions.

Summarizing, using the single-site trial wavefunction, Eq.~\ref{mean_field_trial_function:eqn}, we
find that at small $\Gamma\approx 0.1$ the system is very unlikely to stay in its magnetized phase 
(i.e., the challenging glassy state). 
The lack of regularity in the variational minimization, consequence of the multiplicity of metastable minima, 
makes $\psi_T^{(MF)}$ not a particularly suitable guiding wavefunction for a GFMC simulation.

%----------------------------------------------------------------------------------------------
\begin{figure}[!ht]
\begin{center}
\includegraphics[width=8.0cm]{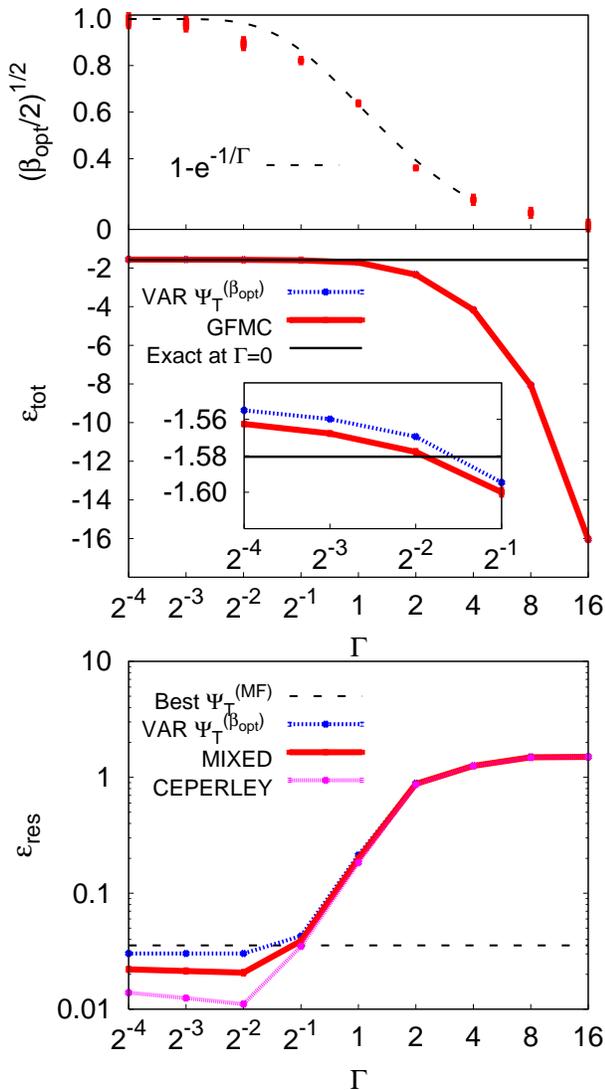}
\end{center}
\caption{(Color online) 
(Top) 
Plot of the optimal $\beta$, $\beta_{opt}$, for the ``Boltzmann" trial wavefunction
$|\psi_{T}^{(\beta)}\rangle$ defined in Eq.~\protect\ref{thermal_trial_function:eqn}, 
for several values of $\Gamma$. The dashed line is a guide to the eye.
(Center) 
Optimal variational energies 
$\epsilon_{tot}^{(Boltz)}=\langle \psi_T^{(\beta_{opt})}|H| \psi_T^{(\beta_{opt})}\rangle/N$ 
corresponding to the $\beta_{opt}$ shown in the Top panel, 
and the GFMC estimate of the total energy per spin $\langle H \rangle/N$, versus $\Gamma$.
The inset magnifies the small-$\Gamma$ region, where small differences are noticeable.
(Bottom) 
The variation residual diagonal energy 
$\epsilon_{res}^{(Boltz)}=\langle\psi_{T}^{(\beta_{opt})}|H_{cl}|\psi_{T}^{(\beta_{opt})}\rangle/N -\epsilon_{GS}$
corresponding to the $\beta_{opt}$ shown in the Top panel, 
and two GFMC estimators of the residual diagonal energy 
$\epsilon_{res}=\langle H_{cl} \rangle/N-\epsilon_{GS}$: The mixed average, and the Ceperley correction (see text).
The dashed horizontal line represents the best residual energy ever achieved, for
$\Gamma>0.01$, by employing the mean-field trial wavefunction in Eq.~\ref{mean_field_trial_function:eqn}.
}
\label{fig_var_gfmc:fig}
\end{figure}
%----------------------------------------------------------------------------------------------
%
A second, quite natural, choice of trial wavefunction is a Boltzmann-like wavefunction of the form:
\begin{equation} \label{thermal_trial_function:eqn}
\psi_{T}^{(\beta)}(\{S_{i}\}) = \mathcal{N(\beta)}\,e^{-(\beta/2)\,E_{cl}(\{S_{i}\})} \;,
\end{equation}
where $1/\beta$ plays the role of an effective temperature, with $\beta$ a variational parameter
to be optimized, and $E_{cl}(\{S_i\})$ (see Eq.~\ref{Ising_glass_hamiltonian:eqn}) 
is the classical energy of a given configuration $\{S_i\}$. 
$\mathcal{N(\beta)}$ is an appropriate normalization factor, which we will not
need to calculate. 
Once again, for large $\Gamma$ we expect to find $\beta=0$
(the exact $\Gamma=+\infty$ solution), while, by decreasing $\Gamma$, larger and larger
values of $\beta$ will favor regions where the ``potential energy'' $E_{cl}(\{S_i\})$ has a local minimum,
until we get, for $\Gamma=0$, to the asymptotic limit $\beta \to \infty$ (ideally), required by a wavefunction 
which is perfectly localized in the global minimum (see below for a discussion of this point). 

To calculate the expectation value of energy with the Boltzmann-like choice 
in Eq.~\ref{thermal_trial_function:eqn}, as a function of the single parameter $\beta$, 
we used a standard Variational Monte Carlo (VMC) algorithm.\cite{Mitas:review}
Fig.~\ref{fig_var_gfmc:fig} shows (top panel) the optimal value $\beta_{opt}$ of $\beta$ which minimizes 
the variational energy $E_{tot}^{(Boltz)}=\langle \psi_T^{(\beta)}|H| \psi_T^{(\beta)}\rangle$, for several 
values of the transverse field $\Gamma$.
Notice that $\beta_{opt}$ saturates for small $\Gamma$ to about $\beta_{opt}\approx 2$, somewhat surprisingly
at first sight, since, for $\Gamma\to 0$, one would expect $\beta_{opt}\to+\infty$, in such a way
that the classical ground state dominates
(i.e. Eq.~\ref{thermal_trial_function:eqn} becomes a delta-like function localized in the exact classical
ground-state).
This is clearly an effect of severe ergodicity loss of the VMC algorithm, which is not
difficult to understand. 
For a given $\beta$, indeed, the $VMC$ samples a probability distribution 
$|\psi_T^{(\beta)}(x)|^2=e^{-\beta E_{cl}(x)}$ with single spin-flip moves: its efficiency in exploring
the phase space, therefore, is exactly identical to that of a classical Metropolis Monte Carlo at
the temperature $T=1/\beta$. 
Finding an optimal $\beta$ for a given $\Gamma$ is therefore totally equivalent to ask what 
is the \emph{effective temperature} of a classical Ising spin glass which provides the best approximation
to the wavefunction of a quantum Ising glass at zero temperature and non-zero $\Gamma$.
Now, from classical spin-glass physics \cite{PMV:book,Castellani-Cavagna:review} 
we know that a threshold energy $E_{th}$ exists below which the system has a finite 
complexity, i.e., it displays an exponentially ($\sim \exp{N}$) large number of metastable minima. 
Close to this threshold energy, the relaxation of any local algorithm towards equilibrium
becomes exceedingly slow (the algorithm gets stuck for a long time in every minimum visited) 
and the average quantities measured are not representative of their true thermodynamical values. 
Evidently, for $\Gamma \to 0$, the variational algorithm is not visiting the regions near the 
true minima of the classical energy, but is wandering in a high-energy band of metastables states,
separated by moderate energy barriers.
In such a case, a small and finite value of $\beta$ allows to still overcome such
barriers, so as to find slightly more favorable local minima, while
perfect localization ($\beta\to +\infty$) in a wrong exited state would lead to an average 
bigger residual energy, 

The central and bottom panels in Fig.~\ref{fig_var_gfmc:fig} show the optimal variational energies 
$\epsilon_{tot}^{(Boltz)}=\langle \psi_T^{(\beta_{opt})}|H| \psi_T^{(\beta_{opt})}\rangle/N$,
and the variational residual energy 
$\epsilon_{res}^{(Boltz)}=\langle\psi_{T}^{(\beta_{opt})}|H_{cl}|\psi_{T}^{(\beta_{opt})}\rangle/N -\epsilon_{GS}$
corresponding to the optimal $\beta$ shown in the top panel, for several values of transverse field $\Gamma$.
For large $\Gamma$ values, the variational total energy (center panel) is linear in $\Gamma$, as it should, 
since the transverse field kinetic term dominates in the quantum paramagnetic phase
(see Eq.~\ref{Ising_glass_hamiltonian:eqn}), while the variational residual energy per-site is of order 1. 
By decreasing $\Gamma$, we notice that the variational residual energy saturates, for small $\Gamma$, 
to finite non-zero values, of order $0.03$, in agreement with the previously noted saturation 
in the optimal $\beta_{opt}$, due to ergodicity breaking.
A closer inspection shows that the variational residual energy is actually {\em non monotonic}
for $\Gamma<0.25$, again an artifact of sampling difficulties.
Notice, however, that this saturation value is definitely below 
the best (down to $\Gamma=0.01$) results provided by the previously discussed $\psi_{T}^{(MF)}$, 
which is of order $0.035$ (shown for comparison by a dashed horizontal line).
Therefore, with all its pitfalls, the Boltzmann-like trial wavefunction in Eq.~\ref{thermal_trial_function:eqn}
provides, at low $\Gamma$, a marginally better approximation of the true GS, than that obtained by the 
mean-field \emph{Ansatz}, Eq.~\ref{mean_field_trial_function:eqn}. 
Moreover, $\psi_T^{(\beta)}$ is also much better behaved, and simpler to optimize, as far as 
the minimization problem is concerned.
For these reasons, we have decided to work out our GFMC results using
the Boltzmann-like wavefunction only.
 
%--------------------------------------------------------------------------
\section{Beyond variational: GFMC at fixed transverse field}\label{static_GFMC:sec}
%--------------------------------------------------------------------------
%
The variational results presented in the previous section can, in principle, be improved by GFMC.
Before showing the actual GFMC annealing results, it is instructive to see the performance
of GFMC at a fixed transverse field $\Gamma$. 
We briefly mention the main ingredients used. We used $20$ walkers, and
performed importance sampling using, for each $\Gamma$, the optimized trial Boltzmann-like 
wavefunction $\psi_{T}^{(\beta_{opt})}$ discussed in Sec.~\ref{Variational:sec}.
Branching was performed every $10$ iterations for large values of $\Gamma$. 
For $\Gamma<0.25$, however, weight instabilities are so severe that one needs 
to perform branching at every iteration.
Finally, we made use of a continuous-time approach, sampling directly the probability
of generating an off-diagonal move with a Poisson's process.\cite{trivedi_ceperley:article}
In Fig.~\ref{fig_var_gfmc:fig} we plot the GFMC results obtained for the $80\times 80$ random 
Ising model instance used in Ref.~\onlinecite{Santoro:science}, for several fixed values of the transverse 
field $\Gamma$. 
The middle panel shows the GFMC estimate of the total energy (per site) for several values of $\Gamma$, 
compared to the variational results, $\epsilon_{var}^{(Boltz)}$, discussed in the previous section. 
The inset allows to appreciate the differences between the two results in the small $\Gamma$ region, which 
are invisible on the scale of the main plot.
In the bottom panel we report several data regarding the residual diagonal energy. 
A small technical point is here in order.
GFMC can calculate directly total energy estimates (or averages of other observables which 
commute with $H$) 
%-- $\langle H \rangle_w$ in the notation of appendix \ref{local_energy:subsec} --,
while averages of operators that {\em do not commute} with the Hamiltonian $H$ are 
less straightforward to obtain.\cite{Mitas:review}
In particular, what one can simply evaluate is the so-called {\em mixed average}
\cite{Mitas:review,trivedi_ceperley:article} of the potential energy $H_{cl}$, 
\begin{equation}
\langle H_{cl}\rangle_w = \frac{\langle \psi_{T}|H_{cl}| \psi_{GS}\rangle}{\langle \psi_{T}|\psi_{GS}\rangle} \;,
\end{equation}
This is the estimator labeled ``mixed'' in Fig.~\ref{fig_var_gfmc:fig}. 
The true expectation value we want is, instead, $\langle \psi_{GS}|H_{cl}| \psi_{GS}\rangle$, 
which might be poorly approximated by the mixed average if the trial wavefunction is poor.
A simple partial cure to this drawback, is to include the so-called {\em Ceperley correction}
\cite{trivedi_ceperley:article}: 
$\langle \psi_{GS}|H_{cl}| \psi_{GS}\rangle \approx 2\,\langle H_{cl} \rangle_w-\langle H_{cl}\rangle_T$ 
where $\langle H_{cl}\rangle_T=\langle\psi_{T}|H_{cl}|\psi_{T}\rangle$
is the variational estimate obtained in the previous section.
The results of the latter approximation for the residual energy are labeled ``Ceperley'' in 
Fig.~\ref{fig_var_gfmc:fig}, and are seen to be consistently lower than the mixed average estimates,
which, in turn, are lower than the variational result. 
However, the non-monotonic behavior of the residual energy data for small $\Gamma<0.25$, previously noted for 
the variational results, should ring a bell about the quality of the trial wavefunction, and the 
efficiency of the sampling (i.e. ergodicity), in that region.

Summarizing, we have shown that a fixed-$\Gamma$ GFMC algorithm, while improving a bit
on the variational results, still suffers from the rather unsatisfactory quality of
the importance sampling trial wave-function, and still shows ergodicity problems in
the small-$\Gamma$ region.

%----------------------------------------------------------------------
\section{GFMC Quantum Annealing}\label{annealing_GFMC:sec}
%----------------------------------------------------------------------
%
We finally present the results of a GFMC-based QA approach, where
the transverse field $\Gamma$ is decreased stepwise during the simulation,
while, at the same time, the importance sampling Boltzmann-like wavefunction 
is changed according to the corresponding value of the variational parameter 
$\beta_{opt}(\Gamma)$.
 
As a benchmark, we will compare GFMC-QA outcomes with the Path-Integral Monte Carlo
quantum annealing (PIMC-QA) and classical simulated annealing (CA) results described 
in Refs.~\onlinecite{Santoro:science,Martonak:ising}.
We reduce the coupling $\Gamma$ in Eq.~\ref{Ising_glass_hamiltonian:eqn}
at each Monte Carlo step (MCS) in a linear way: We start from an
initial large enough value of the transverse field, $\Gamma_0=2.5$, and then
we set $\Gamma_n=\Gamma_0(1-n/\tau)$ during the $n$-th MCS ($0\leq n< \tau$).
$\tau$ is the total annealing-time (see Fig.~\ref{gamma:fig}) measured as the 
total number of MCS performed by the algorithm.
We used $M=20$ walkers, and performed branching at each MCS, because the low-$\Gamma$ 
region is affected by severe weight instabilities which would otherwise make the algorithm 
unstable (for the initial, large-$\Gamma$, part of the annealing one could consider branching 
less often, as weights are well under control; this makes a negligible difference).
However, even with this very conservative choice, the weights go sometimes completely wild 
if a $\Gamma$ of order $10^{-7}$ is reached (i.e. for long annealing-times $\tau$): 
As a consequence, we decided to cut-off the $\Gamma_n$ annealing schedule in such a way
that the final $\Gamma$ is $\approx 10^{-4}$ and not $0$. 
This finally guarantees a good weight stability.
(Whenever it was possible to perform annealings with smaller cut-offs on $\Gamma$,
and approximately the same slope $1/\tau$, we checked that the results obtained are not 
very sensitive to the cut-off chosen.)
For each value of $\Gamma_n$ the trial wavefunction used is the Boltzmann-like one, 
defined in Eq.~\ref{thermal_trial_function:eqn}, with a variational parameter 
$\beta_{opt}(\Gamma_n)$, which corresponds to the instantaneous optimal value.
(Practically, we used for $\beta(\Gamma)$ the fitting function shown
in Fig.~\ref{fig_var_gfmc:fig}, upper panel.)

Fig.~\ref{fig_annealing:fig} shows the best residual energy per spin ever reached during the
annealing simulation, for several values of $\tau$, averaged over $10$ independent repetitions 
of the whole annealing process (due to computer time-limitations, a single run is shown for 
the largest, $\tau>10^8$, annealings).
For comparison, the CA and PIMC-QA data obtained in Ref.~\onlinecite{Santoro:science} are also shown. 
Notice first that the $\tau$ axes of the three calculations are completely unrelated: The GFMC
$\tau$ is measured in units in which a MCS is just a {\em single spin-flip}, while MCS for the
CA and PIMC-QA are intended as {\em sweeps of the entire lattice} of $N$ spins 
(including all the $P=20$ Trotter slices, for the PIMC case). 
For this reason, we also present the CA and PIMC-QA data in a shifted time axis where 
$\tau$ is multiplied by a factor $N=80^2$ (rightmost curves). 
Although the GFMC-QA data are strictly below both the CA and the PIMC-QA data, 
on the same per-spin time-unit (i.e., compared to the shifted CA and PIMC-QA data), it is clear 
that the GFMC-slope is still {\em worse} than that of PIMC-QA, and indeed surprisingly similar to CA.
Moreover, the CPU-time needed for a single spin-flip in GFMC is much larger than the 
corresponding single-spin move in CA or PIMC-QA (each GFMC move costs of order $N$ operations).

Let us pause to consider the similarity between the CA and the GFMC-QA slopes that
Fig.~\ref{fig_annealing:fig} suggests.
This similarity must be somehow related to the fact that we have used, as importance
wavefunction for the GFMC, a Boltzmann-like wavefunction, $\psi_T(x) \propto e^{-(\beta/2)E_{cl}(x)}$.
At the level of a plain variational Monte Carlo (VMC), we already pointed out that
such a choice of wavefunction amounts to sampling 
$|\psi_T^{(\beta)}(x)|^2=e^{-\beta E_{cl}(x)}$, and is therefore totally equivalent to a
classical Metropolis Monte Carlo at temperature $T=1/\beta$.  
If, during the GFMC simulation, we neglect the weights associated to the walkers (as well as the
associated branching process), we will be carrying over a completely classical simulation 
where the generated configurations are distributed (see Eq.~\ref{defnewbx}) according to 
\[
p_{x',x} = \frac{\bar{G}_{x',x}}{\bar{b}_x} \propto 
\Gamma \frac{ \psi_T^{(\beta)}(x') }{ \psi_T^{(\beta)}(x) } 
= \Gamma e^{-(\beta_{opt}(\Gamma)/2)[E_{cl}(x')-E_{cl}(x)]} \;.
\]
As a consequence, the Markov process in $x$-space will obey a classical {\em detailed balance}
condition
\begin{equation}\label{detailed_GFMC:eq}
p_{x',x} \, e^{-\beta_{opt}E_{cl}(x)} = p_{x,x'} \, e^{-\beta_{opt}E_{cl}(x')} \;.
\end{equation}
In other words, a GFMC-QA without weights would be just a computationally heavy way of
doing a classical Simulated Annealing with a peculiar form of annealing of the effective 
temperature $\beta_{opt}(\Gamma)$ (notice, in passing, that such an optimal effective 
temperature never gets too low, since $\beta_{opt}$ saturates to around $\beta_{opt}\approx 2$ 
for low $\Gamma$). 
Quantum Mechanics enters, therefore, only through the weights that the GFMC carries over
(and the unavoidable branching process which makes the multiplicative process of weight
updating numerically stable).
Evidently, such a weight updating is in the present disordered case not sufficiently strong and effective
as to make the resulting averages really different from the underlying classical physics
governing the Markov chain in $x$-space, and the resulting GFMC-QA data are rather close
to the CA ones (although they are much more computer-time demanding).

%-----------------------------------------------------------------------------------------------
\section{Discussion and Conclusions} \label{discussion:sec}
%-----------------------------------------------------------------------------------------------
% 
In this paper we have investigated the practical feasibility of a Green's function
Monte Carlo (GFMC) as a tool for performing Quantum Annealing (QA).
As a natural test case, we have concentrated our attention on the two-dimensional
Edwards-Anderson Ising model in transverse field, which was studied in
Refs.~\cite{Santoro:science,Martonak:ising} using PIMC-QA as well as standard CA.

We identified the choice of the importance-sampling trial wavefunction,
a necessary ingredient in any GFMC, as the crucial step -- as well as the weak point --
of a GFMC based QA (GFMC-QA). 
In particular, we found that the simplest mean-field wavefunction (analogous in many
respects to the Weiss theory of ferromagnetism) is completely useless: 
Its optimization requires finding optimal local fields $h_i$ (or magnetizations
$m_i=\tanh(h_i)\in (-1,1)$) describing the single-site wavefunctions, a complex problem 
with many minima which is as difficult as the original optimization problem, i.e., 
finding a classical configuration of spins $S_i=\pm 1$ optimizing the Edwards-Anderson 
classical term $E_{cl}=-\sum_{\langle ij\rangle} J_{ij} S_i S_j$. 
Using, instead, a simpler Boltzmann-like trial-function 
$\psi_T(x) \propto e^{-(\beta/2)\,E_{cl}(x)}$ (with $\beta$ a tunable parameter
which one can optimize variationally), the resulting GFMC is feasible, but the 
corresponding residual energy results are disappointingly close 
-- in magnitude and in slope, when considered as a function of the annealing time $\tau$ --
to those found by a standard classical simulated annealing (CA) (computationally much cheaper). 
We can rationalize this finding with the inability of the GFMC algorithm, in the present
disordered context, to properly implement the quantum mechanics inherent in the
weights that the walkers carry with them.

At the technical level, we also found that the choice of the number of walkers
and the length interval between two successive branching events, can strongly affect the 
GFMC-QA performance, notably the stability of the algorithm.

The crucial theoretical question is, therefore: 
How can one find good trial variational wavefunctions which describe well
enough the small-$\Gamma$ glassy phase of an Ising spin-glass?
This is, quite evidently, a highly non-trivial task.
Taking inspiration from the existing literature on quantum models without disorder, 
one might think of introducing pair-correlations into the trial wavefunction 
-- for instance, by means of spin-spin Jastrow factors, either at nearest-neighbor or at
longer range -- as usually done in the framework of correlated lattice models,
\cite{CalandraSorella:article} and of electronic structure calculations.\cite{Mitas:review,Hammond:book}
Unfortunately, for a quantum spin-glass, due to frustration and disorder, the form of such pair-correlations is
far from obvious.
Moreover, whenever a large number of variational parameters in the trial-function
is required, very advanced minimization techniques, such as those discussed in
Ref.~\onlinecite{sorella_hessiano:article}, are mandatory.
This kind of computational schemes, however, have been successfully tested only 
in equilibrium simulations of ordered systems, while our GFMC-QA should cope, 
inherently with a non-equilibrium dynamics in a disordered system, a highly
non-trivial step forward.

We conclude that, at present, without a serious effort in constructing reliable importance
sampling variational wavefunctions for a quantum glass, GFMC-QA is not a true competitor of PIMC-QA.

%--------------------- ACKNOWLEDGEMENTS ---------------------------------
%
\begin{acknowledgments}
This project was sponsored by MIUR through FIRB RBAU017S8R004, FIRB RBAU01LX5H,
PRIN/COFIN2003 and 2004, and by INFM (``Iniziativa trasversale calcolo parallelo'').
We acknowledge very useful discussions with Demian Battaglia, Michele Casula, Saverio Moroni, 
Sandro Sorella, Erio Tosatti, and Osvaldo Zagordi.
\end{acknowledgments}
%
%--------------------------------------------------------------------------------
%
%\bibliography{GFMC_Ising} 	% For use with bibtex

\begin{thebibliography}{28}
\bibitem[{foo()}]{footnote10}
\bibinfo{note}{The branching is just a particular Markov process applied to the
  configurations $(x_j,w_j)$, which leads to new walkers $(x'_j,w'_j)$. Each
  new walker $(x'_j,w'_j)$, with $j=1\cdots M$, will have the same weight
  $w'_j=\bar w=\sum_j w_j/M$ and an arbitrary configuration $x'_j$ picked up
  among the $M$ possible old ones $x_k$, $k=1\cdots M$, with a probability
  $p_k$ proportional to the weight of that configuration, $p_k=w_k/\sum_j w_j$,
  see Ref.~\protect\onlinecite{CalandraSorella:article}.}
\end{thebibliography}
%
%--------------------------------------------------------------------------------
% Manually inserted bibliography
%

%--------------------------------------------------------------------------------
%
\end{document}